# Effect of High Frame Rates on 3D Video Quality of Experience

Amin Banitalebi-Dehkordi[1], *Student Member, IEEE*, Mahsa T. Pourazad[1,2], *Member, IEEE*, and Panos Nasiopoulos[1], *Senior Member, IEEE*

[1]Department of Electrical and Computer Engineering, University of British Columbia, Canada
[2]TELUS Communications Inc. Canada
{dehkordi, pourazad, panosn}@ece.ubc.ca

*Abstract*—In this paper, we study the effect of 3D videos with increased frame rates on the viewers' quality of experience. We performed a series of subjective tests to seek the subjects' preferences among videos of the same scene at four different frame rates: 24, 30, 48, and 60 frames per second (fps). Results revealed that subjects clearly prefer higher frame rates. In particular, Mean Opinion Score (MOS) values associated with the 60 fps 3D videos were 55% greater than MOS values of the 24 fps 3D videos.

## I. INTRODUCTION

With the introduction of 3D TV technology to the consumer market, one of the main concerns of content providers is to ensure that a high 3D quality of experience (QoE) is delivered to the viewers. Meeting consumer expectations remains a challenge as 3D visual quality depends on many factors, among which motion, sharpness, and texture have a significant contribution to the viewers' visual experience. It is well known that 3D scenes with several fast-moving objects may be very unpleasant for viewers. Recently, the film industry has tried to address this issue by introducing higher frame rates (HFR) [1]. Increasing the frame rate reduces motion blur, improves the picture sharpness and has the potential to make the 3D viewing experience less cumbersome as well as reducing the chances of viewers experiencing nausea and headaches. While viewers seem to have liked the 3D movies shot at a higher frame rate than normal (24fps), more systematic studies are needed to explain the effect of high frame rate on the viewers' QoE [2]. In the case of 2D, there has been a lot of research done on studying the effect that frame rates have on the perceived video quality [3-7]. As a result, schemes were designed to adaptively adjust the frame rate when needed in order to improve the overall quality [8-10]. In the case of 3D, however, similar studies are at early stages [11][12].

In this paper, we try to systematically study the relationship between the 3D video frame rates and the perceived quality. To this end, we captured several representative 3D video sequences at different frame rates and performed a series of subjective tests to understand the effect frame rates have on 3D perception. Subjects were asked to rate the quality of the videos according to their preference. We used the Mean Opinion Score (MOS) values resulted from the test to find the relationship between visual quality and frame rate.

## II. CAPTURING CONFIGURATION AND 3D VIDEO DATASET

In order to study the effect of the frame rate on 3D video quality, we chose four different frame rates, 24, 30, 48, and 60 frames per second (fps), as these are the rates implemented in the existing cameras. More specifically, 24 fps is the frame rate of cinema films, 30 fps is the rate used in NTSC TV content delivery, 48 fps is double the film rate and has been recently used in a few 3D movies, and 60fps is double the rate of (NTSC) broadcast content.

In our capturing setup, as Fig. 1 shows, we used four identical HD cameras positioned in parallel. Two of these cameras are used to shoot a side-by-side stereo pair in 60 fps and the other two in 48 fps. Then, by skipping every other frame in each of the captured sequences we generated two new sequences with 30 and 24 fps, respectively.

In order to secure temporal synchronization of the two cameras, a single remote control was employed to activate both of them at the same time instance. The temporal synchronization of the video sequences is also adjusted/confirmed by our post-processing algorithm. We chose different and representative scenarios for our video data set to have scenes with low and high level of motion and depth. Each video sequence is 10 seconds long. Our data set is available at our website [13].

## III. SUBJECTIVE TESTS

Our subjective tests were conducted in accordance to ITU-R BT. 500-13 [14]. Sixteen subjects participated in our subjective study with ages between 19 and 37. They were all screened for color and visual acuity (using Ishihara and Snellen charts), and also for stereo vision (Randot test – graded circle test 100 seconds of arc). The evaluation was performed using a HD 3D TV with passive polarized glasses. The 3D display and the settings were based on the MPEG recommendations for the subjective evaluation of the

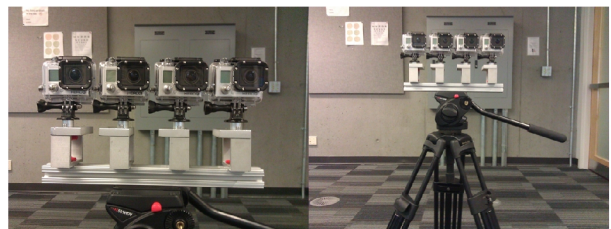

Fig. 1. Camera configuration setup

This work was supported in part by NSERC under Grant CRDPJ 434659-12 & the ICICS/TELUS People & Planet Friendly Home Initiatives at UBC.

proposals submitted in response to the 3D Video Coding Call for Proposals [15].

At the beginning of the experiment, a training session was provided to help viewers become familiar with the test process and show them the expected range of quality-change. Note that the training video was excluded from the subsequent testing process. We used 5 stereoscopic videos with HD resolution (1920x1080) for the test. The test sessions were set up based on the Double-Stimulus Continuous Quality-Scale (DSCQS) method. This method is cyclic in a sense that each time subjects are asked to view a pair of stereo videos from the same scene but with different frame rates and rate the quality of both. In each test session, the test video sequence was shown at 60 fps and also at a lower frame rate in a random order. Afterwards, the subjects were asked to rate both videos without knowing which video is with higher frame rate. For each video, subjects had to rate the quality of the test videos in a continuous scaling format from 0 to 100 (ranges of 0-20, 20-40, 40-60, 60-80, and 80-100 represent "bad", "poor", "fair", "good", and "excellent" video quality, respectively). Once the subjective test results were obtained, we performed an outlier detection analysis. There was one outlier, which was removed from the rest of the data (note that a subject is labeled as outlier if the correlation between the MOS and the subject's rating scores for all videos is less than 0.75).

## IV. RESULTS AND DISCUSSION

After collecting the subjective results, we calculated the Mean Opinion Score (MOS) values for each of the videos. Fig. 2, shows the average difference between the MOS values for the 60 fps 3D videos, and those for the 3D videos at lower frame rates with 95 % confidence intervals (in the DSCQS subjective test, the difference between the MOS of the reference and the impaired video is analyzed [14]). As it can be observed from Fig. 2, the difference between the MOS values for the 60 fps 3D videos and other frame rates is very high (23.41, 29.55, and 55.01 percent improvement in MOS for the 48, 30, and 24 fps videos, respectively). This means that subjects significantly preferred 60 fps over the other rates. It is also worth mentioning that in the case of 60 fps, all of the videos were rated as "excellent" (MOS greater than 80%) and that, with only one exception, all of the videos with 24 fps were rated as "poor" (MOS between 20% and 40%).

In our future work, we are planning to perform extensive subjective tests for wide range of frame rates (1-120 fps) and at the same time determine the best trade-off between QoE and compression/bandwidth. Moreover, the effect of frame rate on 3D QoE can also be incorporated in designing 3D quality metrics.

## V. CONCLUSION

In this paper we studied the relationship between the frame rate of 3D videos and 3D visual quality of experience. We captured several representative stereoscopic videos at different frame rates: 24, 30, 48, and 60 fps. We performed a series of subjective tests to collect assessors' opinions about the visual quality of these videos. These tests showed that higher frame rate 3D videos are clearly preferred to lower frame rate ones. In particular, Mean Opinion Score (MOS) values associated with the 60 fps 3D videos were 55 % greater than MOS values of the cinema 24 fps, with the former ones rated always as excellent.

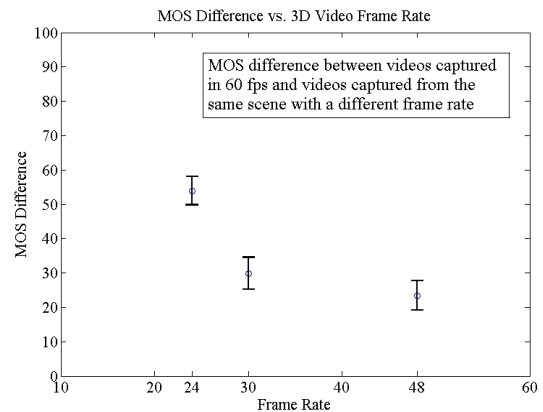

Fig. 2. MOS difference between 60 fps videos and other frame rates-95 % confidence intervals